\documentstyle[12pt]{article}

\def\beq{\begin{equation}}
\def\eeq{\end{equation}}
\input{tcilatex}

\begin{document}

\author{{\large M.Yu.Khlopov{\small $^{1,2,3,4}$}\thanks{%
e-mail: mkhlopov@orc.ru} ,~R.V.Konoplich{\small $^{1,2,3,4}$}\thanks{%
e-mail: konoplic@orc.ru} ,~R.Mignani{\small $^{1}$}\thanks{%
e-mail: Roberto.Mignani@roma1.infn.it} ,~S.G.Rubin{\small $^{2,4}$}\thanks{%
e-mail: sgrubin@orc.ru}} \\
and~A.S.Sakharov{\small $^{2,4}$}\thanks{%
e-mail: sakhas@landau.ac.ru}}
\title{{\huge {\bf Physical Origin, Evolution and Observational Signature of
Diffused Antiworld }}}
\date{{\small {\it $^{1}$}}Dipartimento di Fisica ''E.Amaldi'', III Universita' di
Roma ''Roma Tre'', \ Rome, Italy, and INFN, Sezione di Roma III\\
$^2$Center for CosmoParticle Physics "Cosmion", Moscow, Russia\\
$^3$Institute of Applied Mathematics, Moscow, Russia\\
$^4$Moscow Engineering Physics Institute (Technical University), Moscow,
Russia}
\maketitle

\begin{abstract}
The existence of macroscopic regions with antibaryon excess in the baryon
asymmetric Universe with general baryon excess is the possible consequence
of practically all models of baryosynthesis. Diffusion of matter and
antimatter to the border of antimatter domains defines the minimal scale of
the antimatter domains surviving to the present time. A model of diffused
antiworld is considered, in which the density within the surviving
antimatter domains is too low to form gravitationally bound objects. The
possibility to test this model by measurements of cosmic gamma ray fluxes is
discussed. The expected gamma ray flux is found to be acceptable for modern
cosmic gamma ray detectors and for those planned for the near future.
\end{abstract}

It was recently stated \cite{1}that the existence of antimatter domains with
scale up to 1000 Mpc in a baryon-symmetric Universe cannot escape
contradiction with the observed gamma background. This argument does not
exclude the case when the Universe is globally baryon asymmetric and the
antibaryons within antimatter domains contribute a small fraction of the
total baryon charge. On the other hand, it was shown \cite{2} that the
existence of antimatter domains in the baryon dominated Universe is a
profound signature for the origin and evolution of baryon matter and its
inhomogeneity. Depending on its parameters, the mechanism of inhomogeneous
baryosynthesis can lead to both high- and low-density\ antibaryon domains.
According to \cite{3}, high-density domains can evolve into antimatter
stellar objects, so that a globular cluster of antimatter stars can exist in
our Galaxy, what may be tested in the cosmic searches for antimatter planned
for the near future. On the other hand, as it was mentioned in \cite{3},
low-density antibaryon domains cannot evolve into gravitationally bound
objects. The case of such ''diffused antiworld'' is considered in the
present paper.

A simple modification of the baryogenesis scenario will lead to formation of
domains with different signs of baryon asymmetry. To achieving this, the
condition should be satisfied that C- and CP- violation must have different
signs in different space points . For instance, this can occur if there
exist two sources of CP violation, a spontaneous and an explicit one.
Spontaneous symmetry breaking naturally proceeded through the second order
phase transition which started after the inflationary stage \cite{4}. One
would expect that CP- odd amplitudes are not universal, space-independent
quantities, but have different signs and values in different space points: $%
\phi _{s}=\phi _{s}(x)$. If the amplitude of spontaneous symmetry breaking
is small in comparison with the explicit one, $\phi _{s}<\phi _{e}$ , there
would be relatively small fluctuations in the baryonic number density around
the uniform baryonic background. In the opposite case, $\phi _{e}<\phi _{s}$
, the amplitude of fluctuations is large with respect to the baryonic
background density. One can see that if the domains with high antibaryonic
density are formed in the second-order phase transition, their elementary
size $l_{i}$ at the moment of formation is determined by $l_{i}\simeq
1/(\lambda T_{c})$ \ \cite{5}, where $T_{c}$ is the \ Ginzburg critical
temperature at which the phase transition takes place and $\lambda $ is the
self-interaction coupling constant of the field which breaks CP symmetry. In
this case different elementary domains would expand together with the
Universe and now their size would reach the value $l_{0}\simeq
l_{i}(T_{c}/T_{0})=1/(\lambda T_{0})\simeq 10^{-5}cm/\lambda $, where $T_{0}$
is the present temperature of the background radiation.

Another possible mechanism for generation of large fluctuations of baryon
asymmetry in space is based on the model of baryogenesis with baryonic
charged condensate. At first, the baryon asymmetry is accumulated in the
form of a condensate of the scalar superpartner of the colorless and
electrically neutral combination of quark and lepton fields $\langle
X\rangle $ \cite{6}. This condensate might be formed during the inflationary
stage if baryonic and leptonic charges were not conserved and the potential $%
U(X)$ had a flat direction. The subsequent decay of this condensate would
result in a considerable baryon asymmetry. Depending upon the initial
conditions in the different space points, it is possible to produce baryons
or antibaryons in the decay of the condensate $\langle X\rangle $ \cite{7}.
The characteristic size of an elementary domain with a definite sign of
baryonic charge was estimated in \cite{7}. At the end of inflation it is
equal to $l_{i}\simeq H^{-1}\exp (\lambda _{X}^{-1/2})$, where $H$ is the
Hubble parameter at the end of inflation and $\lambda _{X}$ is the constant
of self-interaction of the field $X$.

The next possibility to generate antimatter domains is connected to the
existence of topological defects of the CP-violating axionic field. The
existence of the vacuum angle $\theta $ leads to strong CP nonconservation 
\cite{4}. The famous solution of the strong CP problem is the introduction
of the Peccei-Quinn (PQ) symmetry together with an axion field \cite{8,9}.
In this solution, $\theta $ has the dynamical meaning of axion field
amplitude, and goes to zero for zero vacuum expectation value of this
amplitude. However, in the early Universe the effective vacuum angle $\theta 
$ varies in the range $[0,2\pi ]$, and this could imply a significant role
of strong CP violation in baryogenesis. The point is that, in the standard
model of invisible axion, one of the consequences of the effective vacuum
angle $\theta $ is the existence of P- and CP- odd condensates. Consider the
cosmological properties of the $N=$ $1$ axion model. Here $N=$ $1$ means
that the axion model has a unique vacuum. The following potential embodies
the qualitative features of the model of interest to us: 
\begin{equation}
L=\frac{1}{2}\partial _{\mu }\phi ^{+}\partial ^{\mu }\phi -\frac{\lambda }{4%
}(\phi ^{+}\phi -F_{a}^{2})^{2}+\Lambda ^{4}(1-\cos \theta )  \label{1}
\end{equation}
where $\phi =$ $\phi _{1}+i\phi _{2}$ is a complex scalar field. For $%
\Lambda =$ $0$ this model has a $U(1)$ global symmetry under which $\phi
(x)\mapsto \exp (i\theta )\phi (x)$. This $U(1)_{PQ}$ symmetry is
spontaneously broken by the vacuum expectation value $\langle \phi \rangle =$
$F_{a}\exp (i\theta )$. The associated Nambu-Goldstone boson is the axion.
The last term in \ Eq.({1)} represents the non-perturbative QCD effects that
give the axion its mass $m_{a}$. The model has global string solutions for $%
\Lambda =$ $0$. A straight global string along the $z$-axis is the static
configuration $\langle \phi \rangle =$ $F_{a}f(\rho )\exp (i\theta )$, where 
$(z,\rho ,\theta )$ are cylindrical coordinates. The surprising result that
emerges from numerical simulations is that most of the energy in the string
network corresponds to infinite strings. Precisely, the density in infinite
strings was estimated to be the $80\%$ of the total string density \cite{10}.

In the early Universe, the standard axion is essentially massless in the
time interval from the PQ phase transition at temperature of order $F_{a}$
to the QCD phase transition at temperature $\simeq 1GeV$. During that epoch,
axion strings are present as topological defects. At first the strings are
frozen in the plasma stretched by the Hubble expansion. However the plasma
becomes more dilute with time and, when the temperature decreases below $%
T\approx 2\cdot 10^{7}GeV(F_{a}/10^{12}GeV)^{2}$ \cite{11}, the strings move
freely and decay efficiently to axions. Since this decay mechanism is very
effective, there is approximately one string per horizon from the
temperature $T$ to the temperature $T_{1}\simeq 1GeV$, when axion acquires
mass. The existence of linear topological defects implies that the effective
angle $\theta $ varies by $2\pi $ around the string, and this fact might be
responsible for spatial modulation of the baryonic charge density in the
Universe. Let us suppose, according to \cite{3}, that, owing to the spatial
variation of the CP -violating axionic phase, the baryon excess is also
space dependent and given by 
\begin{equation}
B(x)=A+b\sin {\theta (x)}.  \label{2}
\end{equation}
Here $A$ is the baryon excess induced by explicit CP-violation processes,
which gives rise to global baryon asymmetry of the Universe, and $b$ is the
measure of the baryon asymmetry dynamically induced by axion. The antibaryon
excess is generated within a wedge, which is symmetrical relative to a sheet
along the direction $\theta =$ $3\pi /2$, and the facet of this wedge is
connected to the axionic string. The intersection of such wedges gives a
domain with antibaryonic excess. Thus the typical size of an elementary
domain will be as large as the horizon scale $l_{i}\simeq m_{Pl}/T_{BS}^{2}$%
, where $m_{Pl}$ is the Plank mass and $T_{BS}$ is the temperature of
baryogenesis. Note that the defect formation can be triggered by inflation 
\cite{12}. If it takes place during the end of the inflation epoch, the
defects will not be completely diluted away but can increase the typical
distance between strings and, consequently, the typical size of an
elementary domain.

In any case all the above-mentioned mechanisms are able to generate domains
of antimatter of any baryon density and on any scale. Such domains may
consist of many elementary domains of small scale, as in the case of the
standard phase transition, or it can appear like one moderately inflated
elementary domain as in the case of the phase transition triggered by
inflation.

Let us consider the case of low-density antimatter regions, corresponding to
the model of diffused antiworld.

When the density of antimatter $\rho _{\bar{b}}$ within a domain is 3 orders
of magnitude less than the baryon density $\rho _{b}$ (which we assume in
the further discussion corresponding to $\Omega _{b}h^{2}=$ $0.1$ , where $h$
is the Hubble parameter normalized to $100kms^{-1}Mpc^{-1}$), the
cosmological nucleosynthesis in the period $t\simeq 1-10s$ results in a
nontrivial chemical composition. For $10^{-4}<\rho _{\bar{b}}/\rho
_{b}<10^{-3}$ antideuterium is the dominant product. For smaller densities
of antibaryons within domains no antinuclei are formed. At the density $\rho
_{\bar{b}}/\rho _{b}<10^{-4}$ , owing to the low antimatter density inside
the domain, no recombination takes place at $z\simeq 1500$ and the
antimatter domain remains ionized after recombination in regions of baryonic
matter. The radiation pressure and the energy density dominance within the
domain suppress then the development of gravitational instability, so that
the antimatter domains of sufficiently large size should be now clouds of
ionized positron - antiproton plasma presumably situated in voids.

Let us give a quantitative estimation of the surviving size and the
observational effects of domains of a diffused antiworld. Let us assume that
the Universe contains regions of very small antibaryon excess density. At
temperatures above several MeV these regions cannot be strongly affected by
the diffusion of surrounding particles, because their mean free path length
is small enough. Therefore we will consider the evolution of these
antibaryon domains at temperatures $4\cdot 10^{3}K<T<10^{9}K$, when the
Universe is radiation -- dominated and contains mainly electrons, protons,
photons and neutrinos. If the size of the antibaryon region is much greater
than the mean free path of surrounding particles we can solve a one --
dimensional problem assuming that the ''initial'' baryon density at $T=$ $%
10^{9}K$ is given by 
\begin{equation}
n_{b}({\bf R},t_{0})=\left\{ 
\begin{array}{ll}
n_{0}, & x<0 \\ 
0, & x>0
\end{array}
\right.  \label{3}
\end{equation}
In this case the diffusion equation for baryons is 
\begin{equation}
\frac{\partial n_{b}}{\partial t}=D(t)\frac{\partial ^{2}n_{b}}{\partial
x^{2}}-\alpha n_{b}  \label{4}
\end{equation}
where $D(t)$ is a diffusion coefficient. The last term in Eq.({4}) takes
into account the expansion of the Universe. Note that, since the antibaryon
component is very small, we neglected it in the diffusion equation.

Let us introduce a new variable $r$ defined as the baryon to photon ratio, $%
r=$ $n_{b}/n_{\gamma }$. Since the evolution of the photon density is given
by the equation $\partial n_{\gamma }/\partial t=$ $-\alpha n_{\gamma }$, we
can rewrite Eq.({4}) in terms of $r$ as 
\begin{equation}
\frac{\partial r}{\partial t}=D(t)\frac{\partial ^{2}r}{\partial x^{2}}
\label{5}
\end{equation}
where 
\[
r({\bf R},t_{0})=\left\{ 
\begin{array}{ll}
r_{0}, & x<0 \\ 
0, & x>0
\end{array}
\right. 
\]
We note that in the problem under consideration photons are distributed
uniformly in baryon domains as well as in the antibaryon ones. Due to
electroneutrality of the media, the motion of protons is strongly related
with the motion of electrons. The last ones interact with photons, and this
interaction defines the diffusion coefficient $D(t)$ for protons. According
to \cite{13} this coefficient (in units where the Boltzmann constant $k_{b}=$
$1$) is given by 
\begin{equation}
D(t)\approx \frac{3T_{\gamma }c}{2\rho _{\gamma }\sigma _{T}}\approx
0.61\cdot 10^{32}Z^{-3}cm^{2}/s,  \label{6}
\end{equation}
where $T_{\gamma }$ and $\rho _{\gamma }$ are the temperature and the energy
density of the radiation, respectively, $c$ is the velocity of light, $%
\sigma _{T}$ is the Thomson cross section, $Z$ is the red shift which is
related with the time $t$ at this stage by $t\approx 2.6\cdot 10^{19}s/Z^{2}$%
. It is convenient to extract the $t$ -- dependence and to write $D(t)$ in
the following form 
\begin{equation}
D(t)=D(t_{0})\left( \frac{t}{t_{0}}\right) ^{3/2}.  \label{7}
\end{equation}
At the initial time $t_{0}$, corresponding to the temperature $T_{0}=$ $%
10^{9}K$, $D(t_{0})$ is given according to Eq.({6}) by 
\begin{equation}
D(t_{0})\approx 1.24\cdot 10^{6}cm^{2}/s.  \label{8}
\end{equation}
Solving Eq.({5}) with the diffusion coefficient ({6}) we find 
\begin{equation}
r({\bf R},t)=\frac{r_{0}}{2}\left\{ 1-\Phi \left[ \sqrt{\frac{5}{8}}\frac{x}{%
\sqrt{D_{0}t_{0}[(t/t_{0})^{5/2}-1]}}\right] \right\}  \label{9}
\end{equation}
where $\Phi $ is the error function.

Assuming $t>>t_{0}$ we get 
\begin{equation}
r({\bf R},t)\approx \frac{r_{0}}{2}\left\{ 1-\Phi \left[ \sqrt{\frac{5}{8}}%
\frac{x}{\sqrt{Dt}}\right] \right\}  \label{10}
\end{equation}
or, in terms of temperature, 
\begin{equation}
r({\bf R},t)\approx \frac{r_{0}}{2}\left\{ 1-\Phi \left[ 0.527\cdot
10^{-4}\left( \frac{T}{T_{0}}\right) ^{5/2}\frac{x}{cm}\right] \right\} .
\label{11}
\end{equation}

If we choose $T=$ $4000K$ at the end of the radiation - dominated epoch,
then 
\begin{equation}
r({\bf R},t)|_{4000k}\approx \frac{r_{0}}{2}\left\{ 1-\Phi \left[ 0.524\frac{%
x}{pc}\right] \right\} .  \label{12}
\end{equation}
It follows from Eq.({12}) that the average displacement of the boundary
between baryon and antibaryon domains is about $\Delta x\simeq 0.2pc$.
Therefore, the primordial antibaryon regions of low density, which grow up
to $1pc$ or more at the end of the radiation -- dominated stage, remain
practically unaffected by the diffusion of the ordinary matter.

Note that according to Eq.({9}) in the limit $t>>t_{0}$ the motion of the
boundary is determined by the final time $t$ and does not depend on $t_{0}$.
This is due to the fact that at the end of the radiation -- dominated stage
the particle number density drops significantly and as a consequence the
mean free-path length increases, thus giving an important contribution to
the boundary motion.

Below $4000K$ atoms are formed in baryon domains. Since the antibaryon
density is assumed to be small enough ($\rho _{\bar{b}}/\rho _{b}<10^{-4}$)
in our approach, we can consider the flow of hydrogen atoms into antibaryon
regions as a motion of free -- streaming atoms. The physical distance
traveled by the atoms after recombination until the present time $t_{p}$ is
given by 
\begin{equation}
d\approx a_{p}\int_{t_{rec}}^{t_{p}}\frac{v(t)dt}{a(t)}  \label{13}
\end{equation}
where $v(t)$ is the average velocity of atoms and $a(t)$ is the scale factor
of the Universe. It is convenient to set $a=$ $ya_{p}$, where $y=$ $1/(1+Z)=$
$=T_{p}/T$ , and to rewrite Eq.({13}) as 
\begin{equation}
d\approx \int_{a_{rec}}^{a_{p}}\frac{v(a)da}{a\dot{a}}%
=\int_{a_{rec}/a_{p}}^{1}\frac{v(y)dy}{y\dot{y}}.  \label{14}
\end{equation}

At the instant of recombination the velocity of atoms is given by the
thermal value $v_{rec}\approx c\sqrt{T_{rec}/m}$, where $m$ is the mass of
the atom. After $4000K$, the typical velocity of atoms $v\approx p/m$ is red
-- shifted down as $1/a\sim T$. Therefore below $4000K$ we have $v\approx
c(T_{p}/m)\sqrt{m/T}(1/y)$. Taking the equation 
\begin{equation}
(\dot{y})^{2}=\frac{8\pi }{3}G\rho (y)y^{2}  \label{15}
\end{equation}
(where $G$ is Newton's constant) and substituting for the density the
expression $\rho (y)=$ $\rho _{0}/y^{3}$ , in the case of a matter --
dominated Universe we find 
\begin{equation}
d\approx \sqrt{\frac{3}{8\pi G\rho _{0}}}\frac{cT_{p}}{m}\sqrt{\frac{m}{%
T_{rec}}}\int_{T_{p}/T_{rec}}^{1}\frac{dy}{y^{3/2}}=\sqrt{\frac{3}{8\pi
G\rho _{0}}}\sqrt{\frac{m}{T_{rec}}}.  \label{16}
\end{equation}
Substituting $\rho _{0}=$ $\rho _{c}$, where $\rho _{c}=$ $1.88\cdot
10^{-29}h^{2}=g/cm^{3}$ is the critical density of the Universe, we obtain $%
d\sim 3/h$ kpc. In the wide range of the Hubble constant $h$ between 0.4 and
1 the free -- streaming length of atoms will be of the order of several kpc.
Therefore antibaryon regions of the same size will be filled at present by
hydrogen atoms.

A key observation to test the model of diffused antiworld could be the
search for gamma rays from a boundary annihilation of antimatter and
hydrogen atoms.

Let us consider first the possibility of annihilation in antimatter domains
filled with hydrogen atoms (the annihilation of matter -- antimatter domains
in baryon symmetric Universe was considered in \cite{1}). The equation for
the number density of antiprotons which takes into account both the
annihilation and the expansion of the Universe is given by 
\begin{equation}
\frac{dn_{\bar{b}}}{dt}=-\langle \sigma v\rangle n_{b}n_{\bar{b}}-\alpha n_{%
\bar{b}}.  \label{17}
\end{equation}

In the limit $n_{b}>>n_{\bar{b}}$ we can neglect the variation of $n_{b}$
due to annihilation. Then , introducing $r=$ $n_{b}/n_{\gamma }$, $\bar{r}=$ 
$n_{\bar{b}}/n_{\gamma }$, and solving Eq.({17}), we find that at present
time in antimatter domains filled with hydrogen atoms 
\begin{equation}
\bar{r}_{p}=\bar{r}_{rec}\exp [-\int_{t_{rec}}^{t_{p}}\left\langle \sigma
v\right\rangle n_{\gamma }rdt].  \label{18}
\end{equation}

Since according to \cite{14} at energies below $10eV$ the cross section of $%
\bar{p}H$ annihilation is given by 
\begin{equation}
\langle \sigma v\rangle \approx 2.7\cdot 10^{-9}cm^{3}/s  \label{19}
\end{equation}
we find that the integral in Eq.({18}) is much greater than unity. So, no
gamma radiation occurs at present from such regions because practically all
antiprotons have been already annihilated.

Therefore the radiation comes only from the narrow region of the boundary
between matter and antimatter domains, where antibaryons have not
annihilated yet. The width of this region is $d\sim v\Delta t$, where $%
\Delta t$ is defined from the condition 
\begin{equation}
\int_{t_{p}-\Delta t}^{t_{p}}\langle \sigma v\rangle n_{b}dt\approx \langle
\sigma v\rangle n_{b}\Delta t\sim 1,  \label{20}
\end{equation}
and the velocity $v$ is given by 
\begin{equation}
v\approx c\frac{T}{m}\sqrt{\frac{3m}{T_{rec}}}  \label{21}
\end{equation}
Substituting all numerical values, we find for the width of the region : 
\begin{equation}
d\approx c\sqrt{3\frac{T_{p}}{T_{rec}}\frac{T_{p}}{m}}\frac{1}{\langle
\sigma v\rangle n_{b}}\approx 0.86\left( \frac{10^{-7}cm^{-3}}{n_{b}}\right)
pc.  \label{22}
\end{equation}
The gamma flux near Earth in this case is given by 
\begin{equation}
\frac{d\Phi }{d\omega d\Omega }\approx \frac{dn}{dt}\frac{dN_{\gamma }}{%
d\omega }\frac{V}{4\pi r_{A}^{2}},  \label{23}
\end{equation}
where $dn/dt=$ $\langle \sigma v\rangle n_{b}n_{\bar{b}}$ is the rate of
annihilation per unit volume and per unit time; $dN_{\gamma }/d\omega $ is
the differential cross section for an inclusive gamma production; $V=$ $4\pi
R^{2}d$ is the volume of the annihilating boundary of the diffused antiworld
at present; $R$ is the size of the diffused antiworld region; $r_{A}$ is the
distance between Earth and the diffused antiworld region.

Integrating Eq.({23}) over photon energy we obtain 
\begin{equation}
\frac{d\Phi }{d\Omega }\approx \langle \sigma v\rangle \beta
n_{b}^{2}\langle N_{\gamma }\rangle d\left( \frac{R}{r_{A}}\right) ^{2},
\label{24}
\end{equation}
where $\langle N_{\gamma }\rangle $ is the average number of photons per one
act of annihilation (typically $\langle N_{\gamma }\rangle \approx 4$) , $%
\beta =$ $n_{\bar{b}}/n_{b}<<1$. Finally, taking into account Eq.({20}) we
get 
\begin{equation}
\frac{d\Phi }{d\Omega }\approx \beta n_{b}\langle N_{\gamma }\rangle \left( 
\frac{R}{r_{A}}\right) ^{2}c\sqrt{3\frac{T_{p}}{T_{rec}}\frac{T_{p}}{m}}%
\approx 2.6\cdot 10^{-4}\beta \left( \frac{n_{b}}{10^{-7}}\right) \left( 
\frac{R}{r_{A}}\right) ^{2}cm^{-2}s^{-1}sr^{-1}.  \label{25}
\end{equation}

The absence of absorption bands in the spectra of long-distance quasars
indicates that the temperature of the intergalactic gas is high enough and
that this gas is ionized. If this is the case we have to change the cross
section in Eq.({19}) by the corresponding cross section for $p\bar{p}$
annihilation \cite{14} 
\begin{equation}
\langle \sigma v\rangle \approx 6.5\cdot 10^{-17}(v/c)cm^{3}/s.  \label{26}
\end{equation}
However, for $p\bar{p}$ annihilation the integral in Eq.({18}) will be much
greater than unity, similarly to the $\bar{p}H$ case. Therefore expression ({%
25}) for the photon flux remains valid for $p\bar{p}$ annihilation because
this flux does not depend explicitly on $\langle \sigma v\rangle $.

Up to now, no evidence for a major anisotropy in the gamma background has
been observed. Below 1 GeV at high latitudes the diffused photon background
is given by \cite{15} 
\begin{equation}
\frac{\Phi _{dif}}{d\Omega }\approx 10^{-6}cm^{-2}s^{-1}sr^{-1}.  \label{27}
\end{equation}

According to \cite{3} a diffused antiworld can exist if $\beta <10^{-4}$.
Therefore it follows from Eq.({25}) that the possible gamma flux from the
annihilation at the boundary of two worlds should be less than $%
10^{-8}cm^{-2}s^{-1}sr^{-1}$. This means that such diffused antiworlds could
exist , in particular not far from our Galaxy, successfully avoiding to be
detected. However, the development of modern detectors of gamma rays, like
EGRET and AMS with the flux sensitivity up to $10^{-8}cm^{-2}s^{-1}sr^{-1}$,
gives a hope to detect diffused antiworlds.

In this article we have found that the minimal scale of an antibaryon
domain, which was not destroyed by annihilation, must be larger then 1 kpc
at present. Independently of the conditions of creation of such an antiworld
domain, the density within it will not increase due to gravitational
instability, provided that two conditions are satisfied. Namely, this domain
must be located in the void and the density of antimatter must be less than
the critical density. It means that the space structure and geometrical form
of the annihilation boundary of such a domain is conserved. In the case when
the antiworld region of scale $R$ consists of many antimatter domains of
minimal annihilation scale $d$, the annihilation in such a region takes
place at the border of each little domain. As a result the annihilation
occurs in the whole volume of this region and, as a consequence, the gamma
flux increases, owing both to the volume effect and the possibility for the
antimatter density within a small domain to be as high as $\beta \approx 1$.
This picture of volume annihilation in antiworld regions can be tested in
searches for the angular variation of gamma flux in the angular range $%
\alpha \sim R/r_{A}$.

\bigskip \bigskip \noindent {\bf Acknowledgments}

\noindent This work was supported in part by the scientific and educational
center ''Cosmion'' and performed within the framework of the International
projects Astrodamus, Coseth and Eurocos-AMS. M.K. and R.K. are grateful to I
Rome University ''La Sapienza'' and III Rome University ''Roma Tre'' for
hospitality and support. We thank B.Kerbikov, A.Kudrjavtsev, and A.Sudarikov
for interesting discussions and suggestions.

\end{document}